# Even odder after twenty-three years: the superconducting order parameter puzzle of Sr$_2$RuO$_4$

A.P. Mackenzie [1,2], T. Scaffidi [3], C.W. Hicks [1] and Y. Maeno [4]

[1] *Max Planck Institute for Chemical Physics of Solids, Nöthnitzer Str. 40, 01187 Dresden, Germany.*

[2] *Scottish Universities Physics Alliance, School of Physics and Astronomy, University of St. Andrews, St. Andrews KY16 9SS, U.K.*

[3] *Department of Physics, University of California, Berkeley, California 94720, USA.*

[4] *Department of Physics, Graduate School of Science, Kyoto University, Kyoto 606-8502, Japan.*

**Abstract**

In this short review, we aim to provide a topical update on the status of efforts to understand the superconductivity of Sr$_2$RuO$_4$. We concentrate on the quest to identify a superconducting order parameter symmetry that is compatible with all the major pieces of experimental knowledge of the material, and highlight some major discrepancies that have become even clearer in recent years. As the pun in the title suggests, we have tried to start the discussion from scratch, making no assumptions even about fundamental issues such as the parity of the superconducting state. We conclude that no consensus is currently achievable in Sr$_2$RuO$_4$, and that the reasons for this go to the heart of how well some of the key probes of unconventional superconductivity are really understood. This is therefore a puzzle that merits continued in-depth study.

# 1. Introduction

The purpose of this short review is to give a status report on research into the superconducting properties, and most specifically the order parameter, of the widely-studied superconductor $Sr_2RuO_4$ [1]. Our approach will be to remain open to all possibilities, and our conclusion will be that the issue is not settled after over twenty years of research. That being the case, it is perhaps worth beginning with a brief discussion of why this is an important problem, worthy of continued research.

Arguably the defining property of a so-called unconventional superconducting state is that the superconducting order parameter has a non-uniform phase in momentum space, such that it can be destroyed by sufficiently strong scattering from non-magnetic disorder [2]. The strength of scattering required depends on the strength of the superconductivity, and $Sr_2RuO_4$ has the most stringent purity criterion for observation of any known superconductor [3]. Its study therefore motivated the growth of extremely high quality single crystals [4], in which it has been possible to determine the Fermi surface and normal state Fermi liquid quasiparticle properties with high accuracy and precision [5–8]. That Fermi surface is relatively simple. It consists of three sheets originating from three $4d$ orbitals of Ru with some contribution from the $2p$ orbitals of oxygen, and is highly two-dimensional. In that sense it is slightly more complicated that the Fermi surfaces of the simplest unconventional superconductors (overdoped cuprates and some organic superconductors), but considerably simpler that those of many heavy fermion or pnictide superconductors. It has therefore been amenable to the construction of accurate but tractable tight-binding models, allowing the a host of modern many-body calculations to be compared with the properties of a real material [9–18].

When one looks at the increasing sophistication of the techniques available for the study of unconventional superconductors, one has the feeling that the $Sr_2RuO_4$ problem really ought to be soluble, for several reasons. Firstly, the normal state is a well-understood Fermi liquid. Secondly, the extremely high purity of the best available samples means that disorder is not nearly as big a complicating factor in experiments as it is in most other materials. Thirdly, the disorder sensitivity of the superconductivity comes because as well as the order parameter being unconventional, the coherence length in the superconducting state is rather long: approximately 750 Å. This means that the thermodynamic features expected of a mean-field, BCS-like transition are seen, and that the superconducting state averages over microscopic detail in a way that is seldom the case for materials with unconventional order parameters whose coherence volumes contain only a few electrons.

The fact that full understanding of the superconducting state of $Sr_2RuO_4$ has not yet been achieved shows the level of challenge that still exists at the interface between theory and experiment in quantum materials, and strongly motivates a new generation of research on this fascinating material. Our goal is to frame that research by highlighting the main problems with finding a fully self-consistent description of the key experimentally determined features of the superconducting state, and to speculate about how the current mysteries might, in future, be resolved. Because a number of lengthy and detailed reviews of the properties of $Sr_2RuO_4$ already exist [3,6,19–22], we will not attempt to be comprehensive. Instead, we will select the issues that we believe to be the most important, and highlight those. We believe that the discussion we will give has significance far beyond understanding the physics only of $Sr_2RuO_4$, because the experimental techniques whose results seem to be in contradiction are among the most commonly used across the whole field of unconventional superconductivity. It is, therefore, worrying that such significant discrepancies exist when they are applied to study some of the best single crystalline samples available of any unconventional superconductor.

## 2. Summary of the theoretical situation

Although the bulk of this paper will be concerned with a discussion of experiment and its interpretation, it is useful to first give some background to that discussion with some general remarks on the current theoretical status of the field.

Soon after the discovery of the superconductivity of $Sr_2RuO_4$ [1], Rice and Sigrist [23] noted that the fact that its normal state is a Fermi liquid with Landau parameters [3,5–7] similar to those of $^3$He [24], and suggested that the analogy might extend to $Sr_2RuO_4$ having a spin triplet superconducting state with an order parameter corresponding to a two-dimensional version of one or more of those seen in superfluid $^3$He. This highly influential paper stimulated interest in the possibility of triplet superconductivity $Sr_2RuO_4$, something that has subsequently been investigated in a long series of calculations by many different groups, each involving different starting assumptions and approximations [10–15,17,25–35]. Later, the importance of spin-orbit coupling was highlighted by a number of authors [36–39], who stressed that the effects of this coupling are strongly **k**-dependent. This means that it might be misleading to use language such as 'spin-triplet' or 'spin-singlet' as descriptors of different superconducting states; a safer language, which we adopt here, is odd or even parity, which does not rely on decoupling the spin and orbital degrees of freedom[#].

The differences between the theoretical predictions that have been made concerning $Sr_2RuO_4$ (which to some extent depend on the input assumptions made) are arguably less important than the common features that have emerged. The most striking of these is illustrated in Fig. 1a for a calculation based on the model of Ref. [33]: Spin-fluctuation theories based on realistic parameterizations of the experimental Fermi surface and mass renormalisations of $Sr_2RuO_4$ find that the free energy difference between odd and even parity states is small. Depending on the input parameters, either parity can be favoured, and among the richer odd parity states, there are also a number of near degeneracies. We stress this point because it immediately illustrates why determining the order parameter symmetry of $Sr_2RuO_4$ is not a trivial problem. Its physical origin almost certainly lies in the structure of χ(**q**,ω). Although difficult to measure with precision, the similarity of the electronic structure of $Sr_2RuO_4$ to that of the itinerant ferromagnets $SrRuO_3$ and $Sr_4Ru_3O_{10}$ and the strongly enhanced metamagnet $Sr_3Ru_2O_7$ indicates the likelihood of enhanced susceptibility near $q$ = 0, a conclusion strengthened by the experiments showing that one of its Fermi surface sheets comes close to van Hove singularities at the M point of the two-dimensional Brillouin zone [40–42]. Some broad weight is seen at low $q$ in inelastic neutron scattering, but those experiments also famously established the existence of a prominent feature at approximately **q** = $(2\pi/3a, 2\pi/3a)$ [43,44]. As might be expected of such an electronic structure [9], the addition of significant levels of dopants such as Ti can stabilize static order at finite $q$ [45] . Crudely speaking, a susceptibility with this kind of $q$ structure can be exploited by many different flavours of spin-fluctuation mediated pairing, so it naturally places $Sr_2RuO_4$ close to the border between odd and even parity superconducting states.

---

[#] In the absence of spin-orbit coupling a pure spin triplet superconductor is odd parity, and a pure spin singlet superconductor is even parity.

The second notable feature, illustrated in Figs 1b and 1c, is the complexity of the predicted gap structures. Even the relatively simple Fermi surface of $Sr_2RuO_4$ introduces considerable variation in the average gap magnitude both between sheets and within a single sheet. Odd parity states may or may not have symmetry-imposed gap nodes, but the ones without nodes have deep gap minima, and the even parity states have a far richer nodal structure than one's naïve expectation based on experience of single-band superconductors.

Depending on one's point of view, Nature is either being unkind or kind here – unkind because of the near degeneracies among different order parameters and the complexity of the gap structures associated with those order parameters make the problem unexpectedly hard, or kind because it offers the prospect of rich superconducting phase diagrams, possibly including transitions between odd and even parity states.

## 3. Identification of key experiments

It is clear from the above discussion that unambiguous determination of the order parameter symmetry of $Sr_2RuO_4$ is likely to require accurate and precise experimental information, because there is not a sufficiently clear difference between the free energies of different candidate states for theory alone to provide a definitive answer. However, the calculations provide guidance on the classes of experiment that are likely to be the most important. Examination of Fig. 1 immediately suggests that thermodynamic data are likely to be complicated, showing signatures beyond those expected of a single gap [46], and this is seen in experiment (Fig. 2). Even qualitative analysis of the temperature dependent heat capacity gives evidence for two or more gaps differing in magnitude by only of order a factor of two [47,48]. The second thing that Fig. 1 suggests is that measurements sensitive to the density of states in the vicinity of gap nodes will need to be performed under extremely stringent conditions if they are to yield definitive information. Ideally, they will need to go to extremely low temperatures (50 mK or below), be performed on the highest purity samples and have the capability of distinguishing accidental nodes or deep gap minima from those imposed by symmetry [49]. Considerable detail and very low temperature measurement will likely be required in order to distinguish one candidate order parameter from another.

The situation outlined above highlights the importance of measurements that are directly sensitive to symmetry. Admirable attempts have been made to conduct parity-sensitive tunneling studies of $Sr_2RuO_4$ [50,51]; while these have generally favoured odd parity superconducting states, the reproducibility from sample to sample is not as good as one would wish, so the results are better regarded as being suggestive than conclusive. There has also been an intriguing observation consistent with the existence of half flux quantum vortices in certain special conditions, again interpreted in terms of an odd parity state with a two-component order parameter, but not yet representing conclusive proof of such a state [52]. Another approach, still in its infancy but holding considerable promise, is the study of the proximity effect between $Sr_2RuO_4$ and metallic magnets [53], for which the predicted behavior for odd and even parity states is substantially different [54].

*Experiments probing time reversal symmetry breaking*

Considerable experimental effort has gone into an explicitly symmetry-related issue, namely investigating whether time reversal symmetry (TRS) is broken on entry into the superconducting state of $Sr_2RuO_4$. Two of the probes most often employed to search for TRS breaking in unconventional superconductors have given a positive result. Muon spin rotation (μSR) indicates the development of spontaneous magnetism near the muon implantation sites even when samples are cooled in zero external field [55], and this conclusion was confirmed in later measurements of magneto-optic polar Kerr rotation [56]. Both the key data sets are shown in Fig. 3. It is not easy to perform a quantitative interpretation of the magnitude of the signal seen in either of these experiments, so the strength and origin of the TRS breaking is not firmly established ♠. Further evidence has been reported from μSR experiments on samples whose $T_c$ is changed by incorporating different levels of non-magnetic impurities that the TRS-breaking signals are associated with the onset of superconductivity [60], but it would be desirable to see more experimental work on this issue.

In spite of the above-mentioned caveats, the prevailing inference from the μSR and Kerr rotation experiments is that the observations result from the order parameter having two degenerate components in its 'orbital' degree of freedom ♣. If this is true, there are important consequences for the likely parity of the superconducting state, because not all candidate order parameter components

---

♠ In this context we note that there a large quantitative discrepancy between the size of the 0.5 G volume-averaged internal fields seen in the muon spin rotation measurements [55] and the much lower limit (≤ 1 mG) on internal fields established by scanning SQUID measurements [57–59].

♣ The possibility that the TRS-breaking might be in the 'spin' degree of freedom has not been widely investigated, though note the caveat above about the difficulties of even using this language in the presence of strongly **k**-dependent spin-orbit coupling effects. Tutorial-style descriptions of how to deduce the symmetry-breaking properties of different odd parity order parameters in the absence of spin-orbit coupling can be found in refs. [3,20]

*are* degenerate in the absence of externally applied fields. The potential significance of this statement can be illustrated by considering the case of a material without spin-orbit coupling. In the tetragonal crystal field of $Sr_2RuO_4$ the only non-accidental way to have two degenerate *d*-wave order parameter components involving intra-band pairing is for them to be $d_{xz}$ and $d_{yz}$ [31], but a TRS-breaking order parameter of the form $d_{xz} \pm id_{yz}$ would feature horizontal line nodes and Cooper pairs formed between electrons in different Ru-O planes [§]. Although not impossible, and indeed also discussed theoretically in the context of odd parity order parameters [12,34], interplane pairing would be a truly exotic state that seems intuitively unlikely in a material with such a strongly two-dimensional Fermi surface . In contrast, *p*-wave components remain degenerate in a tetragonal crystal field, which is why a state of the form $p_x \pm ip_y$ with in-plane pairing has been so extensively discussed in the literature.

One of the expectations of a simple $p_x \pm ip_y$ state is the existence of edge currents which would produce measurable edge magnetic fields. Extensive experimental searches for these edge fields have yielded mostly null results [57–59,61], but in the meantime more sophisticated calculations have suggested a variety of ways in which the edge currents could be far smaller that those predicted by the first naïve estimates [62–65]. More work will be needed to settle this issue completely, but for now it seems as if the lack of observed edge currents does not rule out the existence of a TRS breaking superconducting order parameter in $Sr_2RuO_4$. Another consequence of a two-component order parameter might be the formation of domains in the superconducting state (though we note the comments on this in Ref. [66]). A number of observations are qualitatively consistent with such a hypothesis [51,56,67,68], therefore seemingly favouring the existence of an odd parity order parameter, but the estimates of the characteristic sizes of such domains vary widely.

---

[§] As stressed throughout this article, spin-orbit coupling *is* important in $Sr_2RuO_4$, so examination of the degeneracy-splitting of even parity order parameters based on in-plane Cooper pairing requires explicit numerical calculation using realistic multi-band models rather than simple estimates regarding a purely orbital part of a spin-orbit separable state. Such calculations confirm that the degeneracy splitting of the even parity states is usually substantial: for example, for the parameters used to produce Fig. 1 from the model of Ref. [33], the predicted $T_c$ of a $d_{xy}$ state is approximately one fifth of that of a $d_{x^2-y^2}$ state. In the presence of interactions (included in the model [33] used to construct Fig. 1), these energetic differences become parameter-dependent, and accidental crossings can occur at which different even parity states involving in-plane become degenerate. One can also construct time-reversal-symmetry-breaking even parity states on the three-sheet Fermi surface of $Sr_2RuO_4$ involving inter-orbital pairing, but at the cost that in such states the intra-orbital pairing amplitude would have to be zero. The accurate statement, therefore, is that TRS-breaking condensates of even parity and in-plane Cooper pairs are not impossible, but would require fine-tuning to particular points in parameter space or the imposition of pairing conditions that both seem unlikely in a real material.

*Cooper pair formation and spin susceptibility in the superconducting state*

One of the predictions for a simple even parity superconductor with weak spin-orbit coupling is a strong drop in its spin susceptibility as the superconducting state is entered. This occurs because the non-magnetic singlet Cooper pairs are removed from the reservoir of conduction electrons whose energy can be lowered by field-induced spin polarization [69]. In a superconductor it is not trivial to isolate the contribution of spin to the magnetic susceptibility in the presence of the orbital diamagnetism, but it can be achieved, in appropriate circumstances, by study and analysis of the nuclear magnetic resonance (NMR) Knight shift or inelastic neutron scattering. Spin-orbit coupling complicates the analysis, but it should still be possible in principle to distinguish between the responses of even and odd parity superconductors. For this reason, study of the Knight shift has become a standard probe of unconventional superconductivity.

Such measurements have been performed extensively on $Sr_2RuO_4$; example data sets are shown in Fig. 4. There is no experimental inconsistency in the reported signals: no drop of the extracted spin susceptibility has been seen in any NMR or neutron scattering measurement on $Sr_2RuO_4$ [70–75], and indeed a small rise has been reported in the most precise measurements to date [72]. All of the available data have been interpreted in terms of odd parity order parameters, though the lack of a dependence of the results on field orientation has necessitated the slightly worrying postulate that the vector order parameter can be rotated in extremely small applied fields of order 20 mT. An extensive discussion of the Knight shift measurements and issues involved in their interpretation can be found in Ref. [20].

*Apparently contradictory results*

If the TRS-breaking and spin susceptibility measurements were the only information available about $Sr_2RuO_4$, there would be little doubt that it has an odd parity order parameter. In reality, however, other work favours different conclusions. One prediction for $p_x \pm ip_y$ (and for $d_{xz} \pm id_{yz}$) is that lifting the tetragonal point-group symmetry of the system through in-plane magnetic field [76] or uniaxial pressure [77] should split the transition temperatures of the $p_x$ and $p_y$ components, yielding a double transition. Independent of microscopic detail, the splitting should be proportional to the strength of the applied symmetry-breaking field [77]. However experiments with both in-plane field [78,79] and uniaxial pressure [80,81] have not revealed such splitting.

Arguably an even more worrying discrepancy is revealed by studies of the superconducting upper critical field. In any superconductor, if the energy cost in maintaining equilibrium diamagnetism in the presence of the applied field becomes too high, the superconductivity is lost. It is well known that in standard spin singlet superconductors, contributions to this energy cost come from both creating the diamagnetic response and from a loss of spin energy in forming the Cooper pairs.

If the dominant energy cost comes from the energy required to expel the field by setting up appropriate screening currents, the critical field is often described as being 'orbitally limited'. As we shall see below, this terminology is confusing in the description of modern superconductors, so we will refer to the effect here as 'bulk diamagnetic orbital limiting'. In most superconductors this orbital limiting is dominant, and this is indeed the case for $Sr_2RuO_4$ with the magnetic field applied parallel to the crystallographic $c$-axis ($H_{//c}$). If, however, the material is strongly type II, allowing efficient flux penetration via vortices, a second class of physics can limit the upper critical field if the Cooper pairs are spin singlets. As discussed above, spin singlet Cooper pairs are non-magnetic objects, so their formation results in a loss of magnetic polarization energy. If this loss is overbalanced by the superconducting condensation energy, the condensate forms and the spin susceptibility drops below $T_c$. As the applied field is raised, however, there comes a point at which the magnetic energy overpowers the condensation energy and the superconductivity is destroyed. This process is known as 'spin limiting' or 'Pauli limiting' [82]. For a fully-gapped spin singlet superconductor at $T=0$, the condition is:

$$\frac{1}{2}\chi_P H^2 = \frac{1}{2}N(0)\Delta^2$$

where $\chi_P$ is the Pauli susceptibility, $H$ the applied field, $N(0)$ the density of states at the Fermi level and $\Delta$ the superconducting energy gap. For this mechanism, the persistent currents giving the diamagnetic response disappear because the spin-related energetics result in the premature destruction of the Cooper pairs and hence of the condensate. In a multi-band material, appropriate averages need to be taken, but within factors of order one the prediction is that the limiting will occur when the applied field in tesla is the same as the transition temperature in kelvin. The Pauli limit is therefore a fundamental limit that can be observed in strongly type II superconductors with even parity order parameters. In simple interpretations, it should be entirely consistent with the information obtained from the Knight shift, because both phenomena are rooted in the competition between superconducting condensation energy and spin polarization energy, as illustrated in the sketch in Fig. 5.

For magnetic fields applied in the *ab* plane ($H_{//ab}$), Sr$_2$RuO$_4$ is strongly type II. Its Fermi surface is so anisotropic that for this direction, the critical field based on bulk diamagnetic orbital limiting is expected to be at least a factor of 50 higher than that seen for the field applied parallel to *c* [3,6]. In fact, small angle scattering studies of the vortex lattice have established the intrinsic anisotropy parameter of the superconducting state, which determines the anisotropy of critical fields based on diamagnetic orbital limiting to be 60 [83]. This would predict a critical field of approximately 4.5 T for $H_{//ab}$. In contrast, the measured value is 1.5 T (Fig. 6a) [84] and the transition at low temperatures is first-order, as expected for Pauli limiting [85] rather than the second-order transition expected for diamagnetic orbital limiting [86-88]. Further, the value of 1.5 T is semi-quantitatively consistent with the value estimated by considering the energetics of Pauli limiting ♦.

This observation seems to be qualitatively at odds with the measurements of the NMR or neutron Knight shift, which give no evidence for a spin-related magnetic energy competing with the superconducting condensation energy (note the discrepancy between the combination of Figs. 4 and 6a and the sketch of Fig. 5). The discrepancy has recently become even starker because of uniaxial pressure studies in which the $T_c$ of $Sr_2RuO_4$ was raised to 3.5 K [42]. This was accompanied by an increase of the critical field for $H_{//c}$ by a factor of 20 from 0.075 T to 1.5 T. Given the anisotropy factor of 60, an enormous critical field would then be predicted for $H_{//ab}$, but instead only a modest rise to 4.5 T was observed (Fig. 6b).

The above observations make it seem certain that some critical field limiting mechanism is operating in $Sr_2RuO_4$. Quantitatively, it is in quite good agreement with the simple predictions of Pauli limiting associated with an even parity order parameter, but it is perhaps too early to jump firmly to that conclusion. For example, as pointed out in Ref. [89], odd parity superconductors could themselves experience critical field limiting at some level in the presence of spin-orbit coupling. In these circumstances it is difficult to decouple a microscopic spin susceptibility from a *microscopic* orbital susceptibility, i.e. a magnetic susceptibility arising from the orbital character of the states at the Fermi surface. The issue of how strong this effect could be in $Sr_2RuO_4$ will be a matter for precise calculation using models based on realistic parameterizations of the electronic structure including this spin-orbit coupling. Even if those calculations successfully accounted for the observed critical fields, however, other things would then need to be understood. In particular, it is urgent to obtain a

---

♦ In a simple weakly coupled superconductor with a uniform gap and no magnetic enhancement, the Pauli limit is that $H_{c2}$ in tesla should be a factor of 1.8 times $T_c$ in kelvin [82]. Given the complexity of the gap and the magnetic susceptibility predicted for $Sr_2RuO_4$, direct numerical comparisons should be performed with extreme caution, so the fact that observed value is of order 1 instead of 1.8 should not be overinterpreted.

theoretical understanding that reconciles critical field limiting of *any* microscopic origin with the fact that no associated reduction in susceptibility is observed in the NMR studies.

**4. Summary and future work**

The tone of the summary that one can give about the current situation is probably dependent on one's mood and one's natural levels of optimism or pessimism. The discrepancies that we have outlined above are not minor issues of detail but major qualitative disagreements between the results of experiments that are among the most prominent probes used in the study of superconductivity. At one level, these disagreements are a cause for depression about the state of the field of $Sr_2RuO_4$ physics (and certainly the source of conflict at conferences and meetings!). They also raise uncomfortable questions about how well we understand many other unconventional superconductors. The issue is unlikely to be the quality of the experimental data. The data that we have focused on in this article are not the result of quick and speculative research on poorly-controlled samples, but stem from multiply-verified experiments conducted over many years on samples whose quality is among the highest available in any unconventional superconductor. Instead, the contradictions observed in $Sr_2RuO_4$ suggest that interpretation of some of the key experiments commonly used in the field of unconventional superconductivity is not yet fully understood.

However, these problems can also be viewed as an opportunity. For all the reasons outlined in the introduction, $Sr_2RuO_4$ is a good superconductor on which to refine our understanding. It is therefore important that efforts to resolve the puzzles that it presents are continued and even stepped up. Although it is dangerous to try to predict the future in too much detail, several productive avenues of future research seem clear:

i) There is much to be learned from the kind of studies pioneered in Ref. [53] of proximity effects between $Sr_2RuO_4$ and other magnetic and non-magnetic metals, with renewed efforts desirable on both the relevant experiment and theory.

ii) The existence of reliable and reproducible tunnel junctions into $Sr_2RuO_4$ would enable the parity-sensitive measurements that are so important to distinguish between putative classes of order parameter. This research would likely receive a major boost if sufficiently pure thin films [90] could be grown reproducibly.

iii) The recent uniaxial pressure experiments of Refs. [42,80] offer the possibility in principle of conducting a wide range of experiments on samples whose superconducting transition temperature can be increased by a factor of 2.3, from 1.5 K to 3.5 K. The tuned samples also have a lower crystal symmetry than those at ambient pressure, so in some senses one has the opportunity to create entirely new materials in these experiments. Given the near-degeneracies of competing order parameters highlighted in section 2, it is also possible that order parameter transitions could exist as a function of uniaxial pressure, giving a rich overall phase diagram.

iv) There is evidence from a sharp drop in the critical current of Pb-Ru-$Sr_2RuO_4$ junctions as the temperature is reduced through 1.5 K for interference between the superconductivity around ruthenium inclusions and that in bulk $Sr_2RuO_4$ [91,92]. There is also evidence for internal degrees of freedom in the superconductivity of $Sr_2RuO_4$ from hysteretic and noisy critical currents in such junctions [50,92,93], and in $Sr_2RuO_4$ microbridges [94]. Further investigation of the origin of these observations is highly desirable.

v) Although the complexity of likely gap structures is a complicating factor, it would of course be highly desirable to have high-precision, low-temperature information that established the gap structure of $Sr_2RuO_4$. Recent normal state quasiparticle interference experiments [95] give the hope that extension to low temperatures might become possible.

vi) It has been widely assumed that, with the best samples having mean free paths of microns and the coherence length being over one

hundred times smaller, true clean-limit study of the phase diagram of $Sr_2RuO_4$ had been achieved. Very recently, however, it was reported in Ref. [96] that this might not be the case, with faint signs of superconductivity persisting to higher than expected applied fields in two extremely pure crystals. The authors of Ref. [96] highlight the similarity of such a situation to observations made on organic superconductors, where they are interpreted in terms of entry to a Fulde-Ferrell-Larkin-Ovchinnokov phase. Whether or not this observation is confirmed and extended, it highlights the ever-present need to strive for still further improvements in sample purity. This aspect of higher purity is likely concerned with point disorder, but extended defects such as dislocations or even macroscopic Ru inclusions also need to be carefully monitored, since there is considerable evidence that $T_c$ is increased in their spatial vicinity [97,98]. The recent observations on externally strained samples lead one to speculate that this $T_c$ enhancement is due to internal strain, so obtaining truly pristine $Sr_2RuO_4$ will also likely require detailed knowledge and control of strain fields.

vii) The discussion in this review has largely focused on experiment, but there is also a clear need for continued work on theory, particularly on models concentrating on incorporating the **k**-dependent effects of spin-orbit coupling [37–39] in a realistic way. Is it possible that the contradictions outlined above only appear to be problems because the theories that are being used to frame the interpretation of the key experiments are still missing something? One obvious deficiency in most current theories based on 'realistic' electronic structure is that they are constructed in two dimensions, and hence ignore dispersion and spin-orbit coupling effects that vary with $\mathbf{k}_z$ [37]. This is especially glaring in light of the strangeness of the properties in in-plane magnetic fields, so it is urgent that these theories be extended to the *z*-direction. Further theoretical work on collective modes [14,99] would also be desirable, as this leads to concrete predictions that can be investigated experimentally. In parallel with this, it would be interesting to continue to investigate the precise conditions required for the existence of topological superconductivity in Sr$_2$RuO$_4$ [100].

Overall, then, we prefer the optimistic point of view. In spite of the contradictions that exist in our current understanding of Sr$_2$RuO$_4$, the next decade of research on this fascinating material looks like being at least as exciting as the past two have been.

**Acknowledgements:** We thank M. Sigrist and A. Ramires for useful discussions on critical field limiting, S. Yonezawa and K. Ishida for discussions on recent developments, and S.A. Kivelson and V. Sunko for critical readings of the manuscript. A.P.M. and C.W.H. acknowledge the support of the Max Planck Society, T.S. acknowledges support from the Emergent Phenomena in Quantum Systems initiative of the Gordon and Betty Moore Foundation and Y.M. acknowledges the support of JSPS KAKENHI Nos. JP15H05852 and JP15K021717.

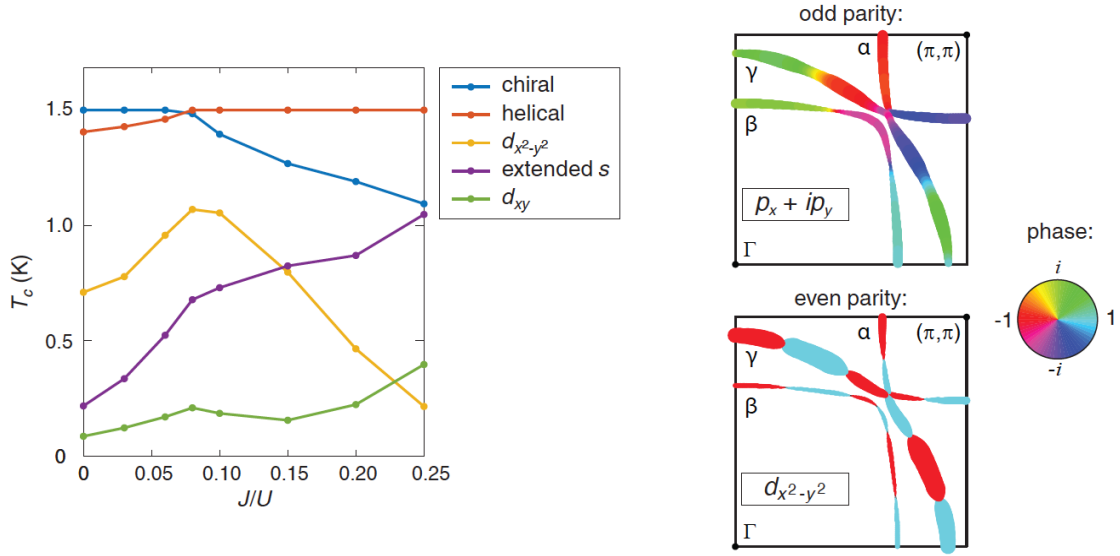

Figure 1: a) $T_c$ calculated for a variety of candidate odd parity (chiral and helical) and even parity states in Sr$_2$RuO$_4$, using the model of Ref.[33] with the maximum calculated value normalised to the experimentally observed 1.5 K. The details clearly have a strong dependence on the model and its underlying assumptions; the data are shown only to illustrate the main point, namely that many different order parameters are close to being degenerate in Sr$_2$RuO$_4$. b) and c) The complicated gap structures predicted for illustrative odd and even order parameter candidates, calculated for $J/U = 0.06$. Panels b) and c) from Ref[42].

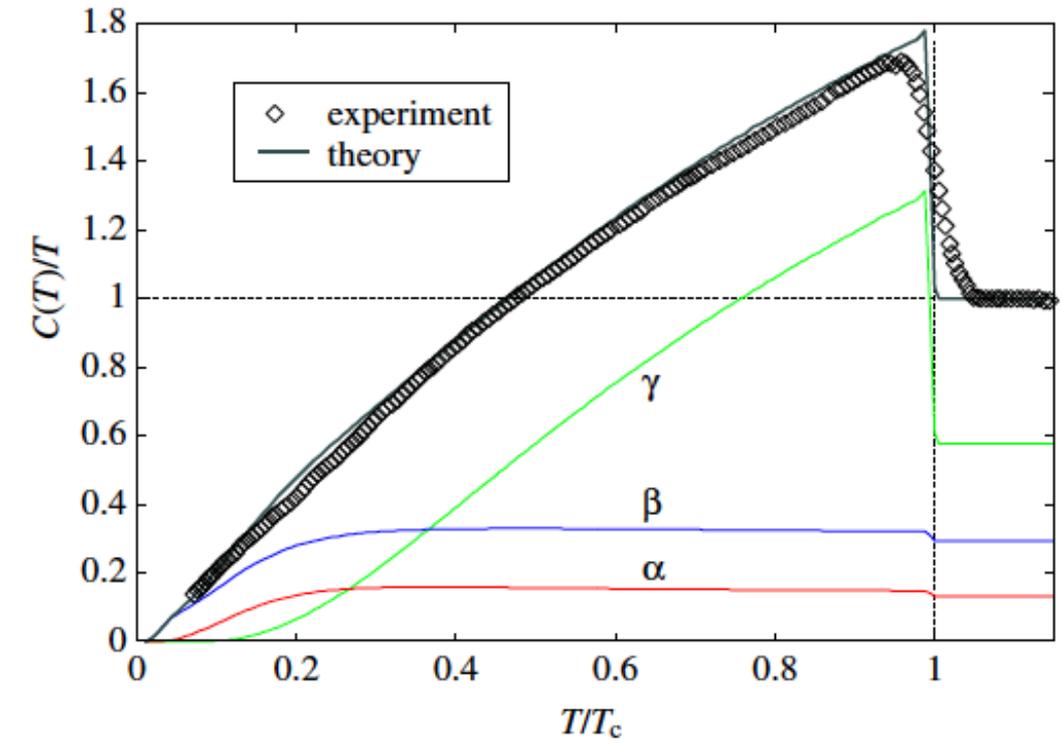

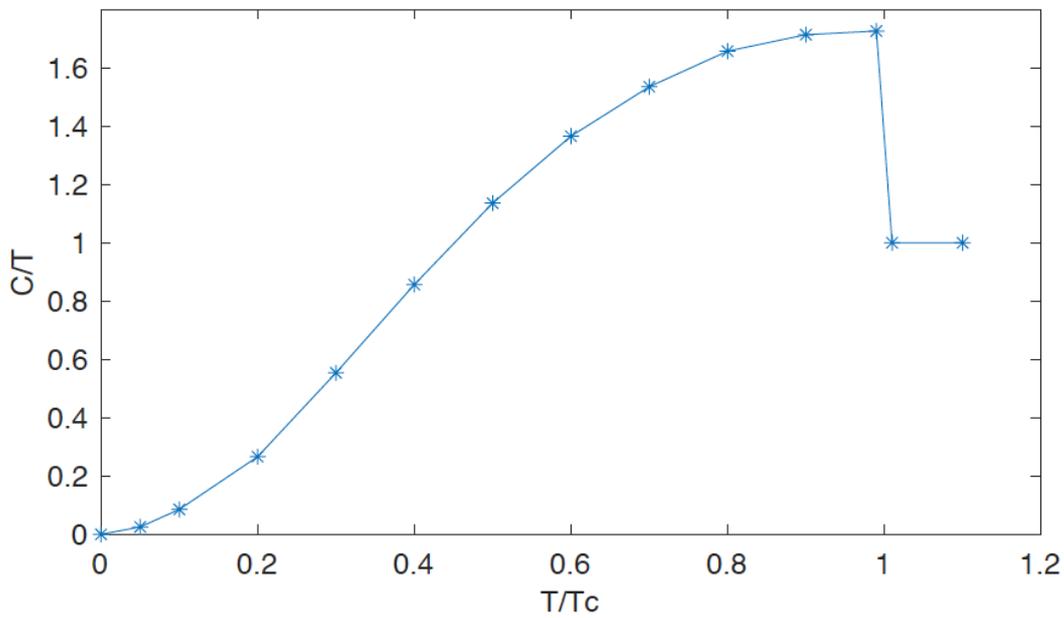

Figure 2: Upper panel - Measured specific heat coefficient of $Sr_2RuO_4$ from Ref. [47] compared with a model calculation from Refs. [10,101], illustrating the decomposition into contributions from the gamma and (alpha + beta) sheets.

Figure from Ref. [101]. Good agreement between experiment and theory (at least for temperatures above 100 mK) is not dependent on details since it only requires that the gaps on the two electronic subsystems be fairly similar in magnitude, and it does not matter whether the (alpha + beta) sheet gap or the gamma sheet gap is the larger one [48]. For completeness, the specific heat prediction for the odd parity gap structure of Fig. 1 b) is shown in the lower panel.

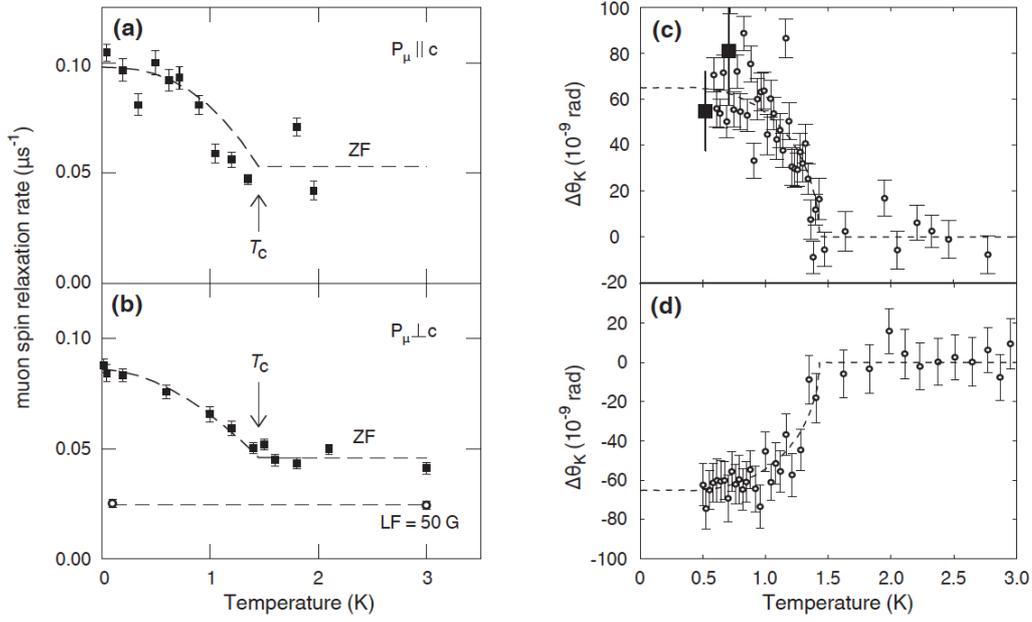

Figure 3: Left panels: the muon relaxation rate for two muon polarisations is seen to increase at the onset of superconductivity in Sr$_2$RuO$_4$, interpreted as signalling the onset of spontaneous fields at defects in the bulk of the sample, due to the superconducting condensate breaking time reversal symmetry (TRS) [55]. Right panels: in a second signal indicative of TRS breaking, a polar Kerr rotation has also been observed at $T_c$ [56]. The sign of the rotation can be trained by cooling in an external field that is then switched off for the warming cycle.

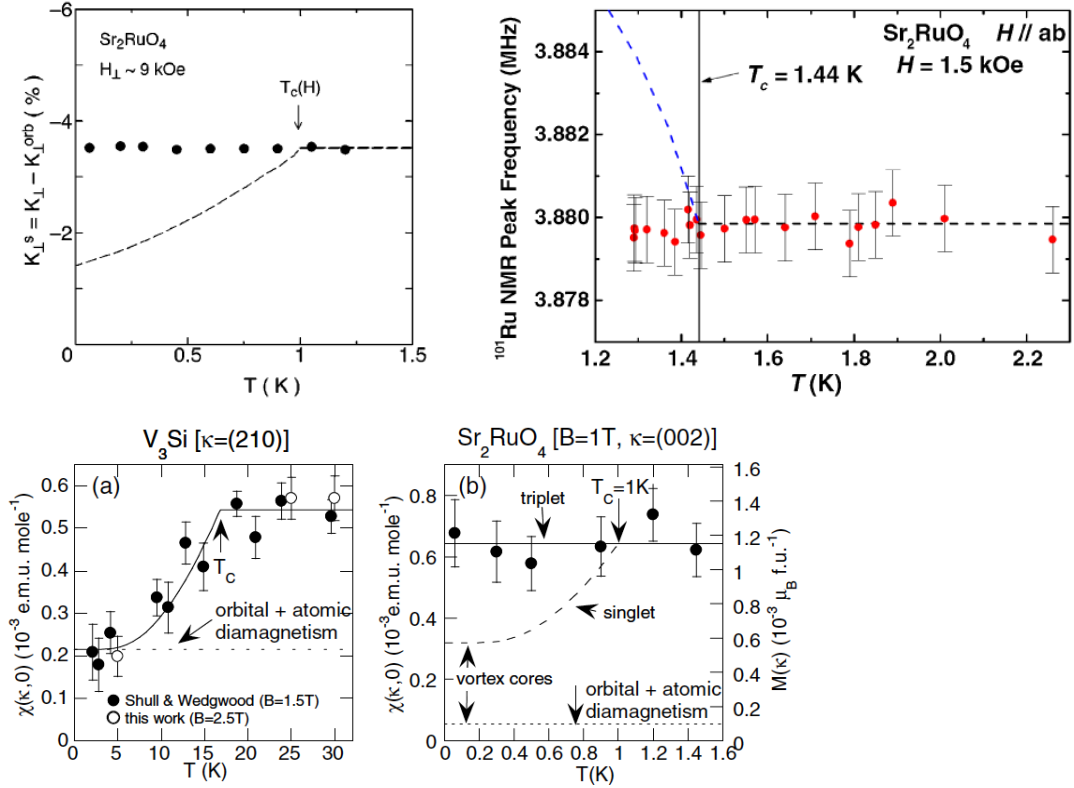

Figure 4: Upper panels: Two examples from the many studies that have been done of the Knight shift in $Sr_2RuO_4$. The examples shown are for $^{99}$Ru nuclei (left) [73] and $^{101}$Ru nuclei (right) [74], for applied fields parallel to the $RuO_2$ planes (note that for $^{101}$Ru, the peak frequency would increase as the spin susceptibility decreased, as sketched by the dotted blue line in the right hand panel). Extensive work has also been done using $^{17}$O nuclei [70] and with variation of the applied field direction. None of the experiments has shown a statistically significant decrease of the Knight shift below $T_c$. Lower panels: The spin susceptibility as deduced from inelastic neutron scattering for the known even parity superconductor $V_3Si$ (left) and $Sr_2RuO_4$ (right). For $V_3Si$ the decrease originally predicted by Yosida [69] is observed, but for $Sr_2RuO_4$ the neutron experiment also resolves no decrease below $T_c$ [71].

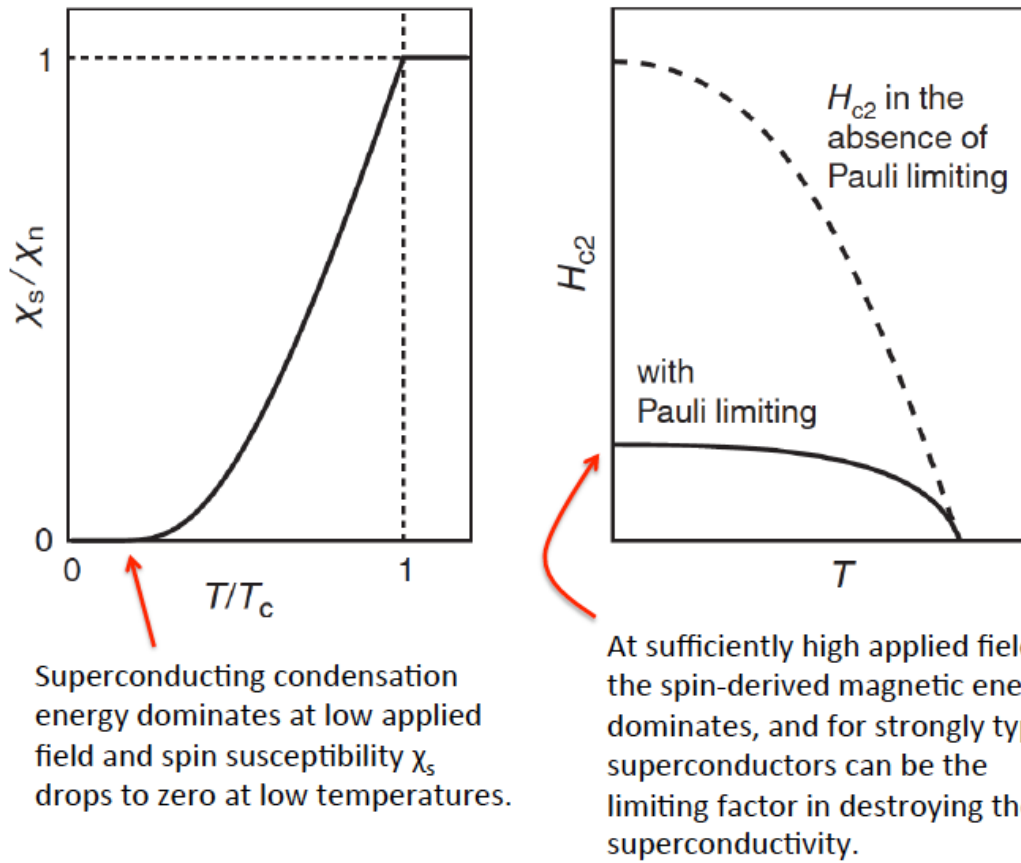

Figure 5: Sketches illustrating the competition between spin polarisation energy and superconducting condensation energy in classic even parity superconductors with spin singlet pairing. Because a singlet Cooper pair is non-magnetic, the magnetic spin polarisation energy gain is lost when the condensate is fully formed at low temperatures. If the applied field is low (left panel) this means that the spin susceptibility is quenched. However, if the applied field is sufficiently high, the spin polarisation energy gain wins out, and the superconductivity is destroyed. Although these sketches are for even parity superconductors in the absence of spin-orbit coupling, and the situation in $Sr_2RuO_4$ is much more complicated, the qualitative relationship between the two measurements would naively be expected to persist. It is thus a major discrepancy that critical field limiting is clearly seen in $Sr_2RuO_4$ but is not accompanied by a decrease of the Knight shift below $T_c$.

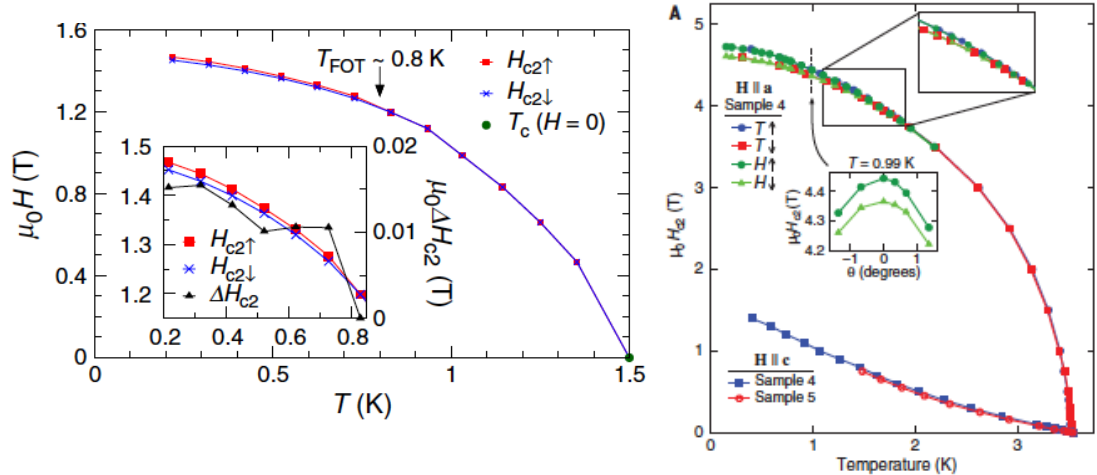

Figure 6: Critical field limiting is seen in both $T_c$ = 1.5 K Sr$_2$RuO$_4$ in ambient conditions (left panel) [86] and in strained $T_c$ = 3.5 K material (right panel) [42]. In both cases there is evidence for a first order transition at low temperatures and high fields. This is a feature of simple theories of paramagnetic limiting in even parity superconductors [85] but might also be expected for magnetic limiting in spin-orbit coupled odd parity superconductors [89].